\documentclass{article}
\usepackage{iclr2027_conference,times}

\iclrfinalcopy          

\usepackage{url}
\usepackage{amsmath}
\usepackage{booktabs, multirow}
\usepackage{xspace}
\usepackage{comment}
\usepackage{xcolor} 
\usepackage{listings} 
\usepackage[most]{tcolorbox} 
\tcbuselibrary{listings,skins} 
\usepackage{inconsolata}
\usepackage{graphicx}
\usepackage{algorithm}
\usepackage{algorithmic}
\usepackage{newfloat}
\usepackage{amssymb}
\usepackage{float}
\newcommand{\ind}{\mathbf{1}}

\usepackage{hyperref}
\definecolor{vscodekeyword}{RGB}{175,0,120}
\definecolor{vscodefunc}{RGB}{70,70,210}
\definecolor{vscodenum}{RGB}{210,90,40}
\definecolor{vscodestring}{RGB}{40,130,40}
\definecolor{vscodecomment}{RGB}{120,120,120}

\lstdefinestyle{python}{
    language=Python,
    basicstyle=\ttfamily\footnotesize,
    keywordstyle=\bfseries\color{vscodekeyword},
    emphstyle=\bfseries\color{vscodefunc},
    commentstyle=\itshape\color{vscodecomment},
    stringstyle=\color{vscodestring},
    emph={len,range,list,str,map,join,extend},
    numbers=left,
    numberstyle=\scriptsize\color{gray!60},
    numbersep=6pt,
    xleftmargin=1.8em,
    frame=none,
    breaklines=true,
    columns=fullflexible,
    keepspaces=true,
    showstringspaces=false,
    aboveskip=0pt,
    belowskip=0pt,
    literate=
        *{0}{{{\color{vscodenum}0}}}{1}
         {1}{{{\color{vscodenum}1}}}{1}
         {2}{{{\color{vscodenum}2}}}{1}
         {3}{{{\color{vscodenum}3}}}{1}
         {4}{{{\color{vscodenum}4}}}{1}
         {5}{{{\color{vscodenum}5}}}{1}
         {6}{{{\color{vscodenum}6}}}{1}
         {7}{{{\color{vscodenum}7}}}{1}
         {8}{{{\color{vscodenum}8}}}{1}
         {9}{{{\color{vscodenum}9}}}{1}
}

\newtcblisting{codestyle}[2][]{
    listing only,
    width=\linewidth,
    colback=blue!7,
    colframe=blue!65!black,
    coltitle=white,
    colbacktitle=blue!55!black,
    title={#2},
    fonttitle=\bfseries\small,
    boxrule=0.7pt,
    arc=1.2mm,
    left=1.5mm, right=1.5mm, top=1mm, bottom=1mm,
    enhanced,
    listing options={style=python},
    #1
}

\newcommand{\slice}{\textbf{SLICE}\xspace}

\floatstyle{ruled}
\newfloat{listing}{tb}{lst}{}
\floatname{listing}{Listing}

\title{SLICE: Specification-Level Isolation of Contract Enforcement}

\author{Soohan Lim, Hyundong Jin, Yo-Sub Han\thanks{Corresponding author.} \\
Department of Computer Science \\
Yonsei University \\
Seoul, Republic of Korea \\
\texttt{\{\protect\href{mailto:aness1219@yonsei.ac.kr}{aness1219},
\protect\href{mailto:tuzi04@yonsei.ac.kr}{tuzi04},
\protect\href{mailto:emmous@yonsei.ac.kr}{emmous}\}@yonsei.ac.kr}}

\begin{document}
\maketitle

\begin{abstract}
Programming problems commonly specify both the computation a function should
perform and the conditions that its inputs must satisfy. Large language models
are widely used to generate code from these problem specifications, and the
generated function must implement the required computation while enforcing the
stated input conditions. The stated input conditions collectively form an
input contract. Enforcing this contract is difficult: incomplete enforcement
accepts inputs that should be rejected, whereas overly restrictive enforcement
rejects inputs that should be accepted. Existing code generation methods do
not provide a generation process that identifies both the input contract and
the functional requirements and generates code that satisfies them jointly.
We therefore introduce \textbf{SLICE}, a generation framework that identifies
both requirements and addresses them through separate generation stages.
SLICE consists of three stages: 
(i) Graph-based specification structuring, which grounds contract conditions
to description segments in a specification graph and removes contract-only
segments to form a functional view;
(ii) Functional body generation, which produces multiple candidate function bodies
through greedy and sampled decoding, ranks them using
execution scores, and resolves ties using difference-region log probabilities; and 
(iii) Contract assertion generation, which generates input-validation
assertions from the identified contract conditions and attaches them to the
selected function body.
We evaluate SLICE on ContractEval across four LLMs and compare it with six
competing methods. Relative to the strongest evaluated baseline for each
model, SLICE improves performance in generating code that satisfies both the
functional requirements and the input contract by an average of 6.58\%.
Our code is available at \url{https://github.com/suhanmen/SLICE}.
\end{abstract}

\section{Introduction}
Programming problems commonly specify both what a function should compute and
which conditions its inputs must satisfy. Code generation benchmarks such as
HumanEval~\citep{chen2021codex}, MBPP~\citep{austin2021mbpp}, and
EvalPlus~\citep{liu2023evalplus} evaluate whether generated programs perform
the required computation correctly, but only on test inputs that already
satisfy the stated conditions. The stated conditions define which inputs the
function is expected to accept and collectively form an input contract. 
An input contract can specify argument types, value ranges, or structural
properties of the inputs~\citep{meyer1992contract}. 
Inputs that satisfy every contract condition are valid inputs, 
whereas inputs that violate at least one contract condition are contract-violating inputs.
A contract-satisfying program must satisfy both parts of the problem specification: it must return correct
outputs for valid inputs and reject contract-violating inputs. 
Because the functional test suites used by these benchmarks contain only valid inputs,
pass@$k$ measures whether the generated program performs the required computation but not whether it enforces the input contract.

ContractEval~\citep{lim2026contracteval} was introduced to evaluate this
missing capability. It pairs each problem specification with two input sets:
valid inputs for evaluating functional correctness and contract-violating
inputs for evaluating contract satisfaction. Because contract conditions are
embedded in natural-language prose alongside functional requirements,
generating contract-satisfying code requires a model to identify these
conditions while implementing the required functionality. Incomplete
enforcement accepts contract-violating inputs, whereas overly restrictive
enforcement rejects valid inputs and reduces functional correctness. Although
ContractEval examines LLM contract enforcement under multiple prompting
strategies, it does not introduce a generation method that jointly preserves
functional correctness and enforces every stated contract condition.

We propose \slice (\textbf{S}pecification-\textbf{L}evel
\textbf{I}solation of \textbf{C}ontract \textbf{E}nforcement), a generation
framework that isolates contract enforcement from functional implementation
through a structured representation of the problem specification. 
SLICE consists of three stages.
(i) Graph-based specification structuring constructs a specification graph
over description segments and extracted contract conditions, 
and removes contract-only segments to construct a functional view.
(ii) Functional body generation produces multiple candidate function bodies
from the functional view through greedy and sampled decoding, ranks them using
execution scores, and resolves ties using difference-region log probabilities.
(iii) Contract assertion generation converts the identified contract conditions
into input-validation assertions and attaches them to the selected function
body. By assigning a distinct objective to each stage, this decomposition
avoids requiring a single generation process to implement the required
functionality and enforce the input contract simultaneously.

We evaluated SLICE on ContractEval using four LLMs. We report functional
correctness~(pass@1) and the Contract Satisfaction Rate~(CSR). We also
introduce the Specification Satisfaction Rate~(SSR), which credits a task only
when the same generated function returns correct outputs for every valid input
and intentionally rejects every contract-violating input. Across the four models, SLICE
improves SSR by an average of 6.58\% relative to the strongest baseline for each model.

We make three main contributions:
\begin{itemize}
\item \textbf{Graph-based specification structuring.} 
We introduce a graph-based method that represents description segments and
extracted contract conditions as nodes. 
SLICE removes contract-only segments while retaining the content required for functional implementation.

\item \textbf{Difference-region candidate selection.}
We propose difference-region log probabilities for selecting among candidate function bodies 
generated through greedy and sampled decoding. By scoring only the
code regions that distinguish candidates with tied execution scores, the method
focuses selection on the implementation choices that differ across candidates.

\item \textbf{Joint specification evaluation and empirical validation.}
We introduce SSR to measure whether the same generated function both achieves
functional correctness and enforces the input contract, and show that SLICE
outperforms all six competing methods in SSR for each of the four LLMs
evaluated on ContractEval.

\end{itemize}


\section{Related Work}

\subsection{Input Contracts and Contract Satisfaction}

Design by Contract defines preconditions, postconditions, and invariants as
executable components of a program specification~\citep{meyer1992contract}.
In function-level code generation, the stated input conditions correspond to
preconditions that determine which inputs the function is expected to accept.
A contract-satisfying function must therefore return correct outputs for valid
inputs and reject contract-violating inputs. 
Standard code generation benchmarks instead evaluate functional correctness 
primarily on valid inputs~\citep{chen2021codex,austin2021mbpp,liu2023evalplus,
hendrycks2021apps,abs-2203-07814,jain2025livecodebench}. HumanEval, MBPP, and
EvalPlus execute generated programs on inputs that satisfy every stated input
condition and report pass@$k$~\citep{chen2021codex,austin2021mbpp,liu2023evalplus}. 
EvalPlus provides program contracts as separate executable
code, but uses them to exclude contract-violating inputs from its functional
test suites rather than to evaluate whether generated programs reject such
inputs. Input contract enforcement therefore remains unmeasured. 
Incomplete enforcement accepts contract-violating inputs, 
whereas overly restrictive enforcement rejects valid inputs and reduces functional correctness.

Existing LLM research on program contracts addresses two related but distinct
settings. One line of work generates contract specifications. Greiner et al.\
generate preconditions and postconditions in the Java Modeling Language~(JML)
from existing Java methods, while Endres et al.\ translate natural-language
intent into postconditions expressed as program
assertions~\citep{greiner2024contracts,endres2024postconditions}. These studies
evaluate the generated contract specifications rather than whether generated
code enforces the input conditions stated in a programming problem. 
Another line of work generates code under explicitly supplied contracts. 
Newcomb et al.\ provide preconditions and postconditions as design constraints for
multi-class code generation and evaluate the resulting implementations through
compilation and functional tests~\citep{newcomb2025preconditions}. 
Because these contracts are supplied separately from the functional requirements
before generation, the model does not need to identify contract conditions
embedded alongside functional requirements in natural-language prose. 
Neither setting directly evaluates whether generated code both achieves functional
correctness on valid inputs and enforces every stated input condition.

ContractEval evaluates this joint requirement by pairing valid tests with
condition-specific contract-violating tests and reporting functional
correctness and contract satisfaction separately~\citep{lim2026contracteval}.
The benchmark establishes the evaluation setting and examines prompting
strategies, but it does not introduce a generation method designed for
contract-satisfying code generation. 
SLICE addresses this gap by representing
the problem specification as a specification graph, identifying contract
conditions before functional body generation, and assigning functional
implementation and contract enforcement to separate generation paths.

\subsection{Code Generation with Large Language Models}

LLM-based code generation has been studied through a range of approaches,
including prompt-based reasoning and
planning~\citep{brown2020fewshot,wei2022cot,jiang2024selfplanning}; execution-guided training,
verification, candidate selection, and
debugging~\citep{chen2024selfdebug,ni2024next,gehring2025rlef};
agent-guided search and multi-agent
generation~\citep{li2025codetree,islam2025codesim,
lee2026solidcoder,xu2026mavencoder}; and specification-oriented
methods~\citep{yan2025codeif,fang2026intentcoding,jia2025specfix}.
CodeTree and SpecFix instantiate agent-guided search and specification repair,
respectively. CodeTree evaluates candidate implementations on provided test
cases and uses their execution outcomes to guide search and
refinement~\citep{li2025codetree}. SpecFix clusters sampled programs according
to their input--output behavior and, when input--output examples are included
in the problem specification, uses them to revise the specification toward
example-consistent behavior~\citep{jia2025specfix}. 
Despite these advances, existing methods are not designed to identify and
enforce the input contract in the problem specification while avoiding
incomplete and overly restrictive enforcement. 
SLICE addresses this gap by extracting contract conditions, and grounding them to
description segments in a specification graph.
It generates multiple candidate function bodies, ranks them using execution
scores, resolves ties using difference-region log probabilities, and generates
input-validation assertions from the identified contract conditions. This
design supports functional correctness while reducing both failure modes.

\begin{figure}[t] 
    \centering
    \includegraphics[width=\textwidth]{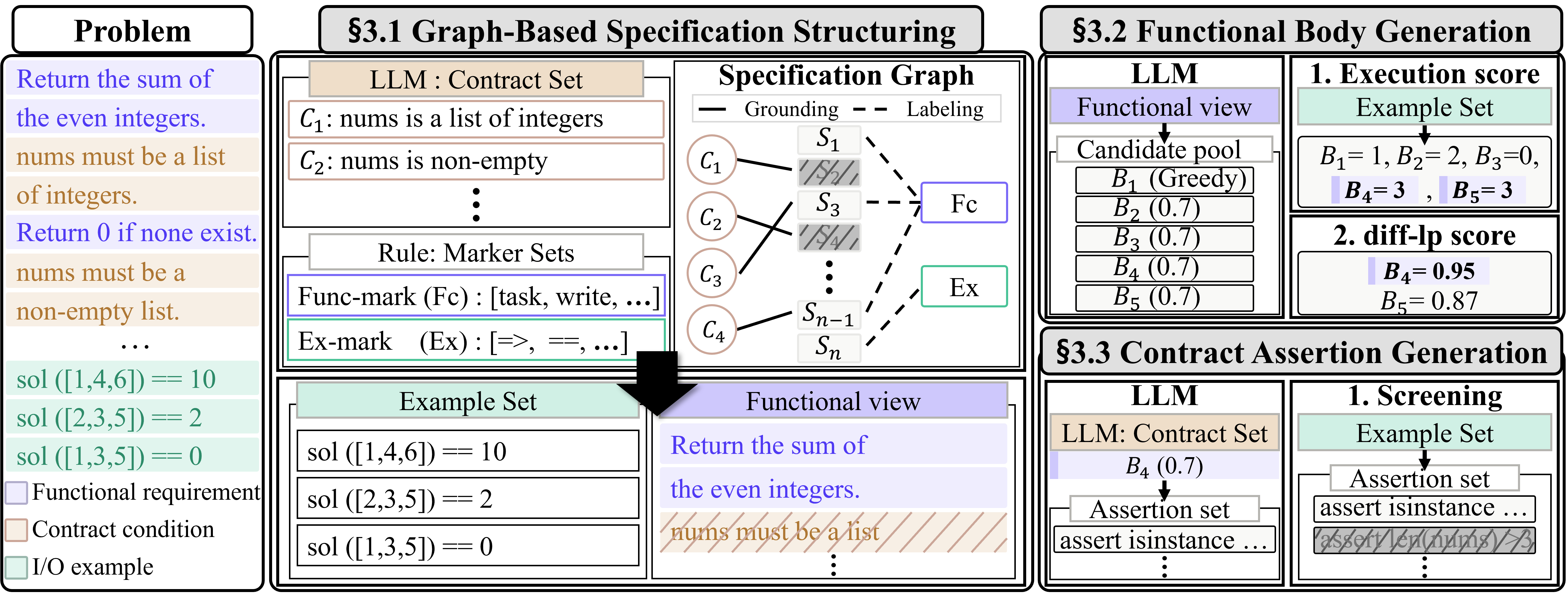}
    \caption{
    An overview of the SLICE pipeline.
    }
    \label{fig:Overview}
\end{figure}


\section{Method}

\slice performs contract-satisfying code generation from a problem specification
without using valid or contract-violating evaluation inputs. A direct
generation prompt requires the model to identify the input contract, implement
the required functionality, and produce input-validation code simultaneously.
This can leave stated conditions unenforced or introduce restrictions that
reject valid inputs. SLICE reduces this burden through three stages with
structured intermediate outputs.
(i) Graph-based specification structuring identifies contract conditions and
constructs a functional view by removing contract-only segments while
preserving functional and example information. The functional view is passed
to Functional Body Generation, the identified conditions to Contract Assertion
Generation, and the preserved example segments to both stages.
(ii) Functional body generation produces greedy and sampled candidates, ranks
them using executable examples, and resolves ties using difference-region log
probabilities. This avoids committing to a single decoding path and focuses
selection on the implementation choices that distinguish tied candidates.
(iii) Contract assertion generation converts the identified conditions into
input-validation assertions using the selected function body as context,
screens them against executable examples, and inserts the retained assertions.
This aligns contract enforcement with the selected functional implementation
while filtering assertions that reject valid examples.
Figure~\ref{fig:Overview} presents the three-stage SLICE pipeline.

\subsection{Graph-Based Specification Structuring}
Generating the function body and input-validation assertions through a single
process can leave stated contract conditions unenforced or introduce
restrictions not stated in the input contract, resulting in incomplete or
overly restrictive enforcement~\citep{lim2026contracteval}. SLICE instead
separates Functional Body Generation from Contract Assertion Generation and
constructs a specification graph that distinguishes contract-only segments
from the content required for functional implementation.

A deterministic parser first divides the problem specification at sentence and
semicolon boundaries, producing a set of description segment nodes denoted by
$\mathcal{D}=\{d_1,d_2,\ldots,d_m\}$. A deterministic labeling rule then
assigns functional and example-preservation labels to each
$d_i\in\mathcal{D}$. A functional label is assigned when $d_i$ contains one
of 15 functional markers, such as \textit{return}, \textit{output}, or \textit{compute}. 
An example-preservation label is assigned when $d_i$ contains one of seven
example-preservation markers, such as \texttt{=>}, \texttt{->}, or \texttt{==}.
These labels define the functional-segment set
$\mathcal{D}_{\mathrm{func}}\subseteq\mathcal{D}$ and the preserved example-segment set
$\mathcal{D}_{\mathrm{ex}}\subseteq\mathcal{D}$, respectively, preserving
segments required for functional implementation and segments used for
execution-based candidate scoring and assertion screening.

SLICE uses the LLM to extract the contract conditions stated in the problem
specification, producing $\mathcal{C}=\{c_1,c_2,\ldots,c_n\}$, where each
$c_j\in\mathcal{C}$ is represented as a contract condition node. For each
$d_i\in\mathcal{D}$, SLICE applies a contract-grounding rule summarized by
the following two criteria.
(i) \emph{Semantic correspondence}: the highest embedding cosine similarity
between $d_i$ and the contract conditions in $\mathcal{C}$ is at least $0.75$.
(ii) \emph{Lexical grounding}: at least $45\%$ of the normalized content words
in $d_i$ also occur in the combined normalized content-word set constructed
from the contract conditions in $\mathcal{C}$ and the function signature.
When the grounding rule is satisfied, SLICE connects $d_i$ to the contract
condition node with the highest embedding cosine similarity, denoted by
$c_{j_i^{\star}}$. Each description segment therefore has at most one
contract-grounding edge. The two criteria serve complementary roles:
semantic correspondence anchors the segment to an extracted contract
condition, while lexical grounding requires contract-related words to account
for a sufficient proportion of the segment. Together, they reduce incorrect
contract grounding, particularly for segments that combine functional and
contract information. Appendix~\ref{ap:graph_construction_rule} lists the marker sets and reports the
accuracy of contract-only masking and executable-example extraction.

SLICE classifies a description segment $d_i\in\mathcal{D}$ as contract-only
when it has a contract-grounding edge and belongs to neither
$\mathcal{D}_{\mathrm{func}}$ nor $\mathcal{D}_{\mathrm{ex}}$. 
A segment in $\mathcal{D}_{\mathrm{func}}$ is retained even when it has a
contract-grounding edge because it also contains information required for
functional implementation. Segments in $\mathcal{D}_{\mathrm{ex}}$ are
retained for execution-based candidate scoring and assertion screening
regardless of their contract-grounding edges. SLICE removes only the
contract-only segments and applies punctuation-level cleanup by removing
dangling connectives and correcting sentence-final punctuation. The resulting
specification forms the functional view, denoted by
$\mathcal{F}_{\mathrm{view}}$, which excludes contract-only content from the
prompt for Functional Body Generation while preserving the functional and
example information required by the subsequent stages.

\subsection{Functional Body Generation}
SLICE generates a fixed candidate pool
$\mathcal{B}=\{b_1,b_2,\ldots,b_5\}$ from
$\mathcal{F}_{\mathrm{view}}$. The pool contains one candidate generated with
greedy decoding and four candidates sampled at temperature $0.7$. Combining
greedy and sampled decoding retains a deterministic candidate while introducing
alternative implementations, allowing SLICE to compare multiple function
bodies rather than commit to a single decoding path. SLICE derives executable
checks from the preserved example segments in
$\mathcal{D}_{\mathrm{ex}}$, forming the check set $\mathcal{E}$. The
execution score of each candidate $b\in\mathcal{B}$ is
\begin{equation}
s_{\mathrm{exec}}(b)
=
\sum_{e \in \mathcal{E}}
\ind\!\left[b \text{ passes } e\right].
\end{equation}
The candidates attaining the maximum execution score form
$\mathcal{B}^{\star}$. If $|\mathcal{B}^{\star}|=1$, SLICE selects the sole
candidate. If $\mathcal{E}=\varnothing$, all candidates receive an execution
score of zero, and $\mathcal{B}^{\star}=\mathcal{B}$.

When $|\mathcal{B}^{\star}|>1$, SLICE resolves the tie using a
difference-region log-probability score, denoted by \texttt{diff-lp}. It
normalizes the tied candidates at the line level and excludes every line shared
by all candidates in $\mathcal{B}^{\star}$, including common signatures,
docstrings, and implementation fragments. For each
$b\in\mathcal{B}^{\star}$, $D_{\mathcal{B}^{\star}}(b)$ denotes the token
positions belonging to lines that are not shared by all tied candidates. Let
$P_{\mathrm{func}}$ denote the token sequence of the complete prompt used for
functional body generation. SLICE concatenates $P_{\mathrm{func}}$ with each
tied candidate and obtains the log probability of each candidate token through
a single forward pass with the same generating model, without generating
additional tokens. When $|D_{\mathcal{B}^{\star}}(b)|>0$, SLICE computes
\begin{equation}
\operatorname{diff\text{-}lp}(b)
=
\frac{1}{|D_{\mathcal{B}^{\star}}(b)|}
\sum_{i \in D_{\mathcal{B}^{\star}}(b)}
\log p_{\theta}
\left(
b_i \mid P_{\mathrm{func}} \Vert b_{<i}
\right),
\end{equation}
where $\Vert$ denotes sequence concatenation, $p_{\theta}$ is the conditional
token distribution of the generating model, and $b_{<i}$ contains the
candidate tokens preceding $b_i$. The causal attention mask ensures that the
probability assigned to $b_i$ depends only on $P_{\mathrm{func}}$ and
$b_{<i}$.

The execution score and \texttt{diff-lp} provide complementary selection
signals. Execution scores favor candidates whose observed behavior matches the
provided examples, whereas \texttt{diff-lp} provides a model-based comparison
when the executable checks do not uniquely identify one candidate. Restricting
the score to differing lines measures the model's support for the
implementation choices that distinguish the tied candidates, while preventing
shared code from dominating the comparison. Averaging over the scored tokens
also prevents longer difference regions from receiving lower scores solely
because they contain more tokens. SLICE selects the candidate in
$\mathcal{B}^{\star}$ with the highest \texttt{diff-lp} score. If all tied
candidates are identical after line-level normalization, SLICE selects the
first candidate according to the fixed candidate order. The resulting
candidate becomes the selected function body passed to Contract Assertion
Generation.
Appendix~\ref{ap:extract_example} reports check coverage and the fallback
behavior, and Appendix~\ref{ap:selection_behavior} analyzes candidate selection.

\subsection{Contract Assertion Generation}

SLICE provides the selected function body and the contract condition set
$\mathcal{C}$ identified during graph-based specification structuring to the
LLM. For each $c_j\in\mathcal{C}$, the LLM generates one input-validation
assertion $a_j$, forming a candidate assertion set $\mathcal{A}$. The contract
condition set restricts assertion generation to the identified input contract,
while the selected function body provides the implementation context needed to
align the assertions with the selected functional implementation. The
candidate assertions are screened before insertion because an assertion
stronger than its source condition can reject inputs permitted by the problem
specification. SLICE derives executable example inputs from the preserved example
segments in $\mathcal{D}_{\mathrm{ex}}$ and treats them as valid because they
are explicitly presented as function examples in the problem specification.
Each assertion is evaluated independently against every available example
input and removed if it rejects at least one input. Expected outputs are not
used because screening checks only whether contract enforcement accepts the
provided valid inputs. The assertions that pass screening form
$\mathcal{A}^{\star}$, and $\mathcal{A}^{\star}=\mathcal{A}$ when no
executable example input is available. SLICE then inserts the assertions in
$\mathcal{A}^{\star}$ at the beginning of the selected function body through a
deterministic AST transformation. The resulting function is the final output
of the SLICE pipeline for contract-satisfying code generation.


\section{Experimental Settings}

\paragraph{Datasets.}
ContractEval~\citep{lim2026contracteval} contains 364 function-level tasks, each
providing a problem specification with an embedded input contract, tests annotated
as valid for evaluating functional correctness, and held-out contract-violating
tests. We excluded 24 tasks with inconsistent annotations, leaving an evaluation
set of 340 tasks. Appendix~\ref{ap:detail_dataset} details the filtering procedure.

\paragraph{Models and baselines.}
We evaluated SLICE on four LLMs: Llama-3.1-8B-Instruct~(Llama-3.1-8B), Qwen3.5-9B,
Gemma-4-12B-it, and GPT-5.4-nano.
We compared SLICE against six baselines: four prompting methods and
two code generation methods. The prompting baselines were direct
prompting~(Base), few-shot prompting~\citep{brown2020fewshot},
chain-of-thought~(CoT) prompting~\citep{wei2022cot}, and
self-planning~\citep{jiang2024selfplanning}. The remaining two were
CodeTree, an agent-guided tree-search method~\citep{li2025codetree}, and
SpecFix, a specification-repair method~\citep{jia2025specfix}.
Appendix~\ref{ap:imple_detail} details the implementation and decoding settings.

\begin{table}[t]
\centering
\footnotesize
\setlength{\tabcolsep}{4pt}
\renewcommand{\arraystretch}{0.92}
\begin{tabular*}{\textwidth}{l@{\extracolsep{\fill}}lccccc}
\toprule
\multirow{2}{*}{\textbf{Model}} & \multirow{2}{*}{\textbf{Method}} & \textbf{Combined} & \multicolumn{3}{c}{\textbf{Individual}} & \textbf{Cost} \\
\cmidrule(lr){3-3} \cmidrule(lr){4-6} \cmidrule(lr){7-7}
& & SSR~($\uparrow$) & pass@1~($\uparrow$) & pass@1$^\star$~($\uparrow$) & CSR~($\uparrow$) & Avg.\ Tokens~($\downarrow$) \\
\midrule
\multirow{7}{*}{Llama-3.1-8B}
& Base            & 46.18\% & 48.53\% & 57.65\% & 94.12\% & 119 \\
& Few-shot        & 44.41\% & 45.00\% & 55.58\% & \textbf{96.18\%} & 173 \\
& CoT             & 43.82\% & 45.88\% & 57.94\% & 95.00\% & 135 \\
& Self-planning   & 43.24\% & 45.30\% & 51.18\% & 93.53\% & 214 \\
& CodeTree        & 49.41\% & 60.00\% & 61.76\% & 82.35\% & 9,225 \\
& SpecFix         & 18.53\% & 50.88\% & 51.18\% & 27.65\% & 19,003 \\
& \textbf{SLICE~(ours)} & \textbf{56.18\%} & \textbf{61.18\%} & \textbf{65.29\%} & 90.00\% & 320 \\
\midrule
\multirow{7}{*}{Qwen3.5-9B}
& Base            & 66.76\% & 68.23\% & 72.05\% & 96.76\% & 201 \\
& Few-shot        & 59.41\% & 59.70\% & 68.82\% & \textbf{97.65\%} & 174 \\
& CoT             & 69.12\% & 71.47\% & 75.88\% & 94.71\% & 238 \\
& Self-planning   & 66.76\% & 68.52\% & 72.34\% & 94.71\% & 235 \\
& CodeTree        & 55.29\% & 61.76\% & 65.00\% & 70.29\% & 15,039 \\
& SpecFix         & 5.29\% & 65.88\% & 66.47\% & 6.47\% & 71,363 \\
& \textbf{SLICE~(ours)} & \textbf{74.41\%} & \textbf{75.88\%} & \textbf{77.94\%} & 95.88\% & 951 \\
\midrule
\multirow{7}{*}{Gemma-4-12B-it}
& Base            & 76.76\% & 79.11\% & 81.76\% & 96.18\% & 287 \\
& Few-shot        & 68.24\% & 68.53\% & 79.41\% & \textbf{98.24\%} & 306 \\
& CoT             & 75.59\% & 77.94\% & 80.59\% & 95.88\% & 311 \\
& Self-planning   & 71.18\% & 74.71\% & 77.06\% & 94.71\% & 289 \\
& CodeTree        & 71.76\% & 80.00\% & \textbf{82.65\%} & 88.82\% & 1,947 \\
& SpecFix         & 28.82\% & 74.41\% & 75.00\% & 33.53\% & 13,054 \\
& \textbf{SLICE~(ours)} & \textbf{77.65\%} & \textbf{80.29\%} & 82.35\% & 95.29\% & 1,056 \\
\midrule
\multirow{7}{*}{GPT-5.4-nano}
& Base            & 70.00\% & 70.29\% & 74.70\% & \textbf{99.12\%} & 155 \\
& Few-shot        & 64.12\% & 64.12\% & 73.24\% & \textbf{99.12\%} & 145 \\
& CoT             & 70.00\% & 71.18\% & 76.47\% & 98.53\% & 159 \\
& Self-planning   & 67.65\% & 70.00\% & 74.41\% & 96.47\% & 272 \\
& CodeTree        & 63.82\% & 66.18\% & 75.88\% & 97.06\% & 2,012 \\
& SpecFix         & 45.00\% & 67.65\% & 68.53\% & 55.29\% & 12,827 \\
& \textbf{SLICE~(ours)} & \textbf{72.65\%} & \textbf{74.41\%} & \textbf{77.94\%} & 97.06\% & 688 \\
\bottomrule
\end{tabular*}
\caption{Four evaluation metrics across four models. Cost is the average
output tokens per task.}
\label{tab:main}
\end{table}

\paragraph{Evaluation metrics.}

We evaluate each generated function using functional correctness~(pass@1),
the Contract Satisfaction Rate~(CSR), and the Specification Satisfaction
Rate~(SSR). For each task, functional correctness requires the generated
function to match the gold code on every valid input, and pass@1 is the
proportion of tasks that satisfy this requirement; pass@1$^\star$ applies the
same criterion after input-validation code is removed. Contract satisfaction
requires the generated function to intentionally reject every
contract-violating input, and CSR is the corresponding proportion.
Pass@1 does not penalize omitted contract enforcement, while CSR alone can be
maximized by rejecting every input. We therefore define SSR as the proportion
of tasks for which both requirements hold for the same generated function:
\begin{equation}
\mathrm{SSR} = \frac{1}{|\mathcal{T}|} \sum_{t \in \mathcal{T}}
\ind\!\left[\, \forall x \in V_t : \hat{f}_t(x) = f_t^{\ast}(x) \,\right]
\cdot \ind\!\left[\, \forall x \in C_t : \mathrm{rej}(\hat{f}_t, x) = 1 \,\right],
\end{equation}
where $\hat{f}_t$ is the function generated for task $t$,
$f_t^{\ast}$ is the gold code, and $V_t$ and $C_t$ are the valid-input
and contract-violating-input sets, respectively.
$\mathrm{rej}(\hat{f}_t, x)=1$ only when $\hat{f}_t$ intentionally rejects
$x$ through its input-validation code. 
By construction, $\mathrm{SSR} \leq \min(\mathrm{pass@1}, \mathrm{CSR})$.
SSR therefore measures whether a generated function both computes the
specified outputs and enforces the input contract, without rewarding
incomplete or overly restrictive enforcement.

\section{Results and Analysis}
The Main Results subsection reports the overall performance of SLICE on
ContractEval in terms of SSR, pass@1, pass@1$^\star$, and CSR. The remaining
subsections examine three complementary aspects of these results. 
The Functional Correctness and Precise Contract Enforcement subsection
examines how all evaluated methods balance body-level functional correctness
and precise contract enforcement, and how this balance translates into SSR.
Generated Token Cost evaluates the
output-token cost of these gains across prompting baselines and prior
code generation methods. Ablation Study examines the contributions of graph-based specification
structuring and candidate selection.

\subsection{Main Results}
Table~\ref{tab:main} reports the performance of SLICE and six competing
methods across the four models. SLICE achieves the highest SSR on all four
models. Relative to the strongest competing method for each model, SLICE
achieves a mean relative SSR improvement of $6.58\%$ across models. The
model-specific improvements are $13.70\%$ on Llama-3.1-8B, $7.65\%$ on
Qwen3.5-9B, $3.79\%$ on GPT-5.4-nano, and $1.16\%$ on Gemma-4-12B-it.
These results show that specification-level isolation consistently improves
SSR, although the strongest competing method differs across models.

The metric-level results show that SLICE achieves this SSR advantage by
preserving functional correctness while maintaining high contract satisfaction.
SLICE records the highest pass@1 on all four models and the highest
pass@1$^\star$ on three; on Gemma-4-12B-it, its pass@1$^\star$ is only
$0.36\%$ below the best result. Across the four models, adding contract
assertions reduces pass@1 relative to pass@1$^\star$ by
$2.50\%$--$6.29\%$, indicating that assertion generation and screening
preserve the functional quality of the selected function bodies. Although
SLICE's CSR is $1.81\%$--$6.43\%$ below the model-specific maximum, the
methods attaining those maxima record lower pass@1 and SSR on every model.
SLICE therefore achieves the highest SSR by combining leading functional
correctness with high contract satisfaction rather than maximizing CSR in
isolation. Appendix~\ref{ap:case_study} presents a task-level comparison that holds
functional correctness constant and isolates contract enforcement.

\subsection{Functional Correctness and Precise Contract Enforcement}

\begin{figure}[t]
    \centering
    \includegraphics[width=0.85\textwidth]{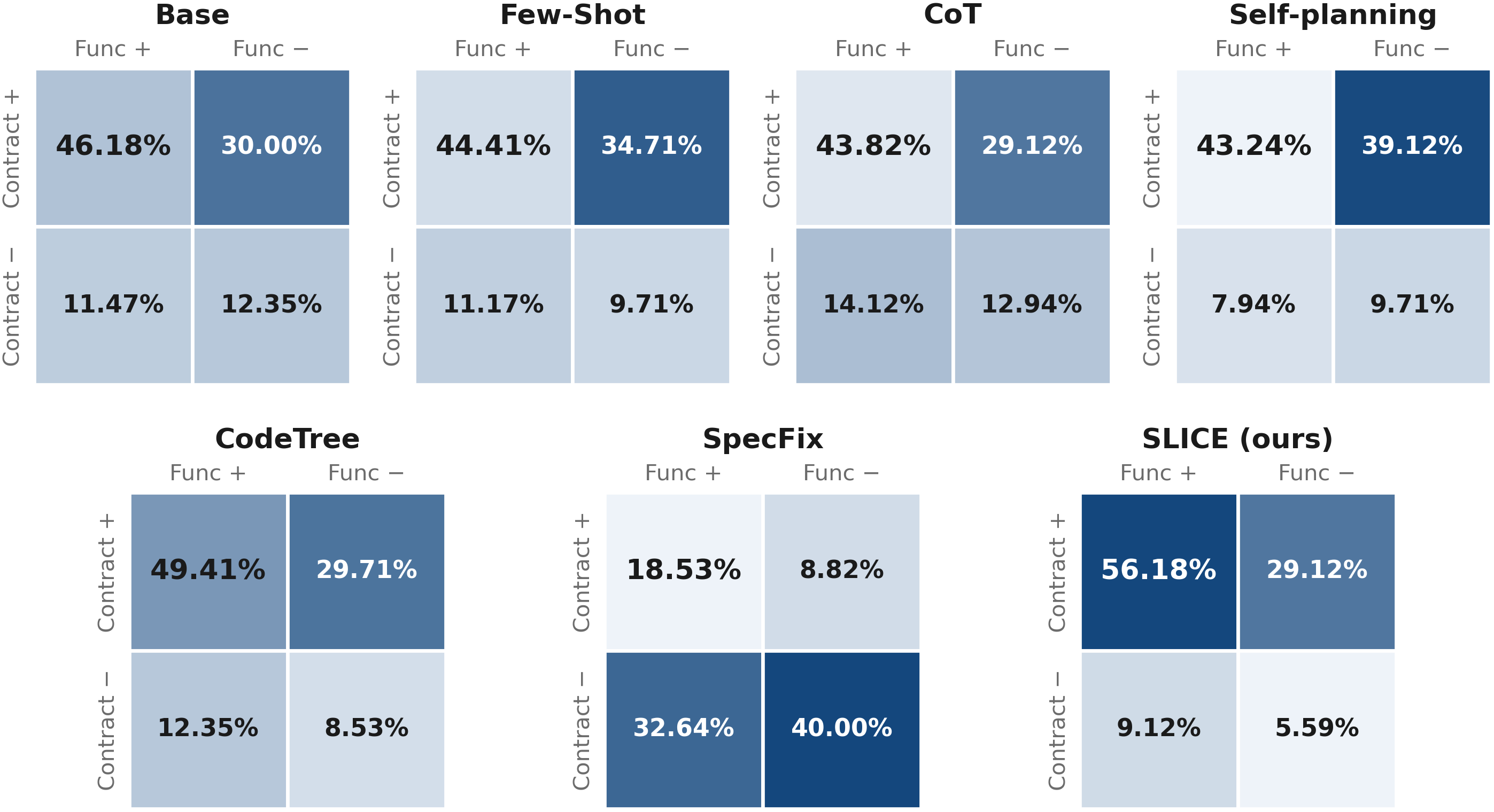}
    \caption{Fine-grained error analysis on Llama-3.1-8B.}
    \label{fig:error_analysis}
\end{figure}

Figure~\ref{fig:error_analysis} decomposes the Llama-3.1-8B results into four
joint outcomes based on body-level functional correctness and precise contract
enforcement. $\mathrm{F}^{+}$ indicates that the implementation passes all
valid tests after its input-validation code is removed, whereas
$\mathrm{F}^{-}$ indicates failure on at least one valid test.
$\mathrm{C}^{+}$ requires the validation code to reject every
contract-violating input without rejecting any valid input, and
$\mathrm{C}^{-}$ denotes failure of either requirement. This criterion is
stricter than CSR, which considers only contract-violating inputs.
Accordingly, $\mathrm{F}^{+}\mathrm{C}^{+}$ equals SSR,
$\mathrm{F}^{-}\mathrm{C}^{-}$ denotes failure in both dimensions, and
$\mathrm{F}^{-}\mathrm{C}^{+}$ denotes an incorrect function body with precise
contract enforcement. $\mathrm{F}^{+}\mathrm{C}^{-}$ denotes a correct body
with contract-enforcement failure and is divided into missed violations and
over-rejection. A task is classified as a missed violation if any
contract-violating input is not rejected; otherwise, it is classified as
over-rejection if any valid input is rejected.
Appendix~\ref{ap:detailed_error} extends the analysis to all four models.

The $\mathrm{F}^{-}\mathrm{C}^{+}$ category isolates functional failure under
precise contract enforcement. Prompting methods assign
29.12\%--39.12\% of tasks to this category, whereas SLICE records
29.12\%, corresponding to a 2.93\% reduction relative to Base, the
strongest prompting baseline. CodeTree records a comparable value of
29.71\%. SpecFix records only 8.82\%, but this value does not indicate
stronger functional implementation when the precise contract-enforcement
criterion holds. Only 27.35\% of its tasks satisfy this criterion, including
18.53\% in joint success and 8.82\% in
$\mathrm{F}^{-}\mathrm{C}^{+}$, while 32.64\% retain a correct body with
precise contract-enforcement failure and 40.00\% fail in both dimensions.
The $\mathrm{F}^{-}\mathrm{C}^{+}$ category must therefore be interpreted
together with the overall rate of precise contract-enforcement success.
The decomposition of $\mathrm{F}^{+}\mathrm{C}^{-}$ shows how each method
fails after obtaining a correct function body. Few-shot records the lowest
contract-violation miss rate at 0.88\%, but over-rejects valid inputs on
10.29\% of tasks. CodeTree exhibits the opposite pattern, with
1.76\% over-rejection and 10.59\% missed violations. SpecFix nearly
eliminates over-rejection at 0.29\%, but misses contract violations on
32.35\% of tasks. SLICE limits the two failure types to 4.12\% and
5.00\%, respectively, reducing the aggregate
$\mathrm{F}^{+}\mathrm{C}^{-}$ rate by 20.49\% relative to Base and
26.15\% relative to CodeTree.
Across all evaluated methods, SLICE assigns the largest share of tasks to
$\mathrm{F}^{+}\mathrm{C}^{+}$ at $56.18\%$ and the smallest share to
$\mathrm{F}^{-}\mathrm{C}^{-}$ at $5.59\%$. Prompting methods record higher
$\mathrm{F}^{-}\mathrm{C}^{+}$ or over-rejection rates than SLICE, CodeTree
misses contract violations on $10.59\%$ of tasks, and SpecFix assigns
$32.35\%$ of tasks to missed violations and $40.00\%$ to joint failure.
Thus, no competing method matches the balance achieved by SLICE across the
complete error distribution. These results show that SLICE improves joint
specification satisfaction by balancing functional implementation and contract
enforcement rather than minimizing one failure type at the expense of another.

\subsection{Generated Token Cost}
Table~\ref{tab:main} reports the average number of generated output tokens per
task. Relative to the highest-SSR prompting baseline for each model, SLICE uses
$168.91\%$--$343.87\%$ more output tokens but improves SSR by
$1.16\%$--$21.65\%$. Its total cost remains between 320 and 1,056 tokens per
task because SLICE performs one contract-condition extraction pass, generates
a fixed pool of five candidate function bodies, and performs one contract
assertion generation pass only for the selected function body; \texttt{diff-lp} produces
no additional output tokens. In comparison, CodeTree and SpecFix require
1,947--15,039 and 12,827--71,363 tokens per task, respectively. SLICE uses
$45.76\%$--$96.53\%$ fewer tokens than CodeTree and
$91.91\%$--$98.67\%$ fewer tokens than SpecFix while achieving higher SSR on
all four models. SLICE therefore converts a bounded increase over prompting
methods into consistent SSR gains and achieves stronger specification
satisfaction than candidate search and specification repair with a
substantially smaller generation budget.

\subsection{Ablation Study} \label{sec:ablation}

Table~\ref{tab:ablation} evaluates two components: graph-based specification
structuring and candidate selection. 
The \emph{$-$ Masking} configuration generates the function body directly from the unmasked problem specification,
leaving contract-only segments mixed with functional requirements. Relative to
SLICE, this configuration reduces SSR and pass@1 by averages of $6.39\%$ and
$7.21\%$, respectively. Its CSR changes by at most $0.61\%$ on Qwen3.5-9B,
Gemma-4-12B-it, and GPT-5.4-nano, but increases by $2.61\%$ on Llama-3.1-8B.
On Llama-3.1-8B, the CSR increase coincides with reductions of $13.47\%$ in
pass@1 and $11.00\%$ in SSR, indicating that retaining contract-only content
can raise contract satisfaction at the cost of correctness on valid inputs, a
pattern consistent with overly restrictive enforcement. Across the four
models, these results show that the graph-derived functional view prevents
contract-only content from interfering with functional implementation without
systematically weakening contract satisfaction. 
The \emph{$-$ Selection} configuration retains graph-based specification structuring, contract assertion
generation, and assertion screening, but replaces the five-candidate generation
and selection procedure with a single greedy body. This change reduces SSR and
pass@1 by averages of $4.07\%$ and $4.36\%$, respectively. CSR remains
unchanged on Llama-3.1-8B and Qwen3.5-9B and decreases by only $0.61\%$ on
Gemma-4-12B-it and $0.31\%$ on GPT-5.4-nano. The performance loss therefore
arises primarily from the functional quality of the selected function body rather than
from contract enforcement. Generating one greedy candidate and four sampled
candidates and selecting among them using executable examples and
\texttt{diff-lp} consistently yields higher SSR and pass@1 than retaining a
single greedy completion, showing that candidate generation and selection
improve functional implementation.

\begin{table}[t]
\centering
\footnotesize
\setlength{\tabcolsep}{4pt}
\renewcommand{\arraystretch}{0.92}
\begin{tabular*}{0.8\textwidth}{l@{\extracolsep{\fill}}lccc}
\toprule
\textbf{Model} & \textbf{Config} & \textbf{SSR}~($\uparrow$) & \textbf{pass@1}~($\uparrow$) & \textbf{CSR}~($\uparrow$) \\
\midrule
\multirow{3}{*}{Llama-3.1-8B}
& SLICE               & \textbf{56.18\%} & \textbf{61.18\%} & 90.00\% \\
& \quad $-$ Masking   & 50.00\% & 52.94\% & \textbf{92.35\%} \\
& \quad $-$ Selection & 51.47\% & 55.00\% & 90.00\% \\
\midrule
\multirow{3}{*}{Qwen3.5-9B}
& SLICE               & \textbf{74.41\%} & \textbf{75.88\%} & \textbf{95.88\%} \\
& \quad $-$ Masking   & 67.94\% & 69.12\% & 95.59\% \\
& \quad $-$ Selection & 71.18\% & 72.35\% & \textbf{95.88\%} \\
\midrule
\multirow{3}{*}{Gemma-4-12B-it}
& SLICE               & \textbf{77.65\%} & \textbf{80.29\%} & 95.29\% \\
& \quad $-$ Masking   & 75.29\% & 77.65\% & \textbf{95.59\%} \\
& \quad $-$ Selection & 76.47\% & 79.41\% & 94.71\% \\
\midrule
\multirow{3}{*}{GPT-5.4-nano}
& SLICE               & \textbf{72.65\%} & \textbf{74.41\%} & \textbf{97.06\%} \\
& \quad $-$ Masking   & 70.59\% & 72.06\% & 96.47\% \\
& \quad $-$ Selection & 71.18\% & 73.24\% & 96.76\% \\
\bottomrule
\end{tabular*}
\caption{Component ablation of SLICE (\%). \emph{$-$ Masking} uses the unmasked
problem specification for body generation. \emph{$-$ Selection} replaces the
five-candidate procedure with a single greedy body.}
\label{tab:ablation}
\end{table}

\section{Limitations} \label{sec:limitation}
SLICE has three limitations. 
(i) \emph{Incomplete separation.} Sentence- and
semicolon-level segmentation avoids fragmenting the problem specification, but
cannot separate functional and contract content expressed within the same
segment. Such segments are retained to preserve functional information,
leaving part of the contract-related content in the functional view.
(ii) \emph{Limited evaluation scope.} Evaluation is limited to ContractEval,
the only benchmark we identified that explicitly provides both valid and
contract-violating tests for this setting. Performance on classes,
multi-function programs, and repository-level code therefore remains unverified. 
(iii) \emph{Higher inference cost.} SLICE performs seven
generation passes per task and additional non-generative forward passes when
\texttt{diff-lp} is required. This design is more expensive than direct
prompting, although its generated-token cost remains lower than that of CodeTree and SpecFix.

\section{Conclusion} \label{sec:conclusion}

We introduced SLICE, a three-stage framework that separates functional
implementation from contract enforcement, and SSR, which measures whether
functional correctness and contract satisfaction hold for the same generated
function. SLICE removes contract-only segments, selects among greedy and
sampled function bodies using executable examples and difference-region log
probabilities, and attaches screened assertions generated from the identified
contract conditions. Across four LLMs on ContractEval, SLICE achieves the
highest SSR for every model. Relative to the strongest competing method for
each model, SLICE achieves a mean relative SSR improvement of $6.58\%$, with
a maximum improvement of $13.70\%$ on Llama-3.1-8B. 
SLICE also achieves these gains at a lower generation cost, using $45.76\%$--$96.53\%$ fewer generated tokens than
CodeTree and $91.91\%$--$98.67\%$ fewer than SpecFix. The ablation results
further show that graph-based specification structuring and candidate
selection both contribute to the performance gains. These results show that
specification-level isolation improves joint specification satisfaction while
preserving functional implementation and enforcing the input contract.

\bibliographystyle{iclr2027_conference}
\bibliography{custom}

\appendix

\section{Implementation Details}\label{ap:imple_detail}
We evaluated the fixed set of 340 ContractEval tasks obtained through the
filtering procedure described in Appendix~\ref{ap:detail_dataset}. The three
open-weight models were evaluated on a machine equipped with three NVIDIA RTX
PRO 6000 GPUs using Hugging Face \texttt{transformers} in
\texttt{bfloat16}. Semantic matching for specification-graph construction
used \texttt{SFR-Embedding-2\_R} with last-token pooling and $L_2$
normalization. Contract-condition extraction and input-validation assertion
generation used greedy decoding with maximum output lengths of 200 and 256
tokens, respectively. Functional body generation produced five candidates per
task: one with greedy decoding and four with sampling at temperature $0.7$ and
top-$p$ sampling with $p=0.95$. All five candidates were capped at 2,048
generated tokens. GPT-5.4-nano was evaluated through its API. Gold code was
used for pre-experiment filtering and post-generation evaluation only.
Neither gold code nor the valid and contract-violating evaluation inputs were
provided to any generation, candidate-selection, or assertion-screening step.

\section{Detailed Dataset Filtering}\label{ap:detail_dataset}
ContractEval contains 364 function-level tasks, comprising 117 HumanEval tasks
and 247 MBPP tasks. We applied two consistency filters before running any
generation experiment. First, we excluded three tasks with no valid-input test.
Because functional correctness is evaluated on valid inputs, these tasks cannot
support a meaningful assessment of whether the generated function implements
the required computation. Second, we excluded 21 tasks for which the gold code
rejected at least one input annotated as valid. These tasks contain a conflict
between the benchmark annotation and the gold-code behavior: a generated
function that reproduces the rejection behavior of the gold code would still
fail the functional criterion because the input is designated as valid.
Retaining these tasks would therefore conflate generation quality with
inconsistencies in the evaluation data.

The two filters removed 24 of the 364 tasks~($6.59\%$), leaving a fixed
evaluation set of 340 tasks~($93.41\%$). The filtering procedure was completed
before any model output was generated and did not depend on the performance of
SLICE or any competing method. All prompting baselines, CodeTree, SpecFix, SLICE, and the ablation
configurations are therefore evaluated on the same task set, with all
task-level metrics computed over the common denominator $|\mathcal{T}|=340$.

\section{Rule-Based Specification Processing}\label{ap:graph_construction_rule}
Table~\ref{tab:marker_sets} lists the marker sets used during
specification-graph construction. Textual markers are matched
case-insensitively, whereas symbolic markers are matched literally. 
Although \texttt{return(s)} appears in both sets, it creates no ambiguity because
functional-marker matching applies anywhere within a description segment
during graph construction, whereas executable-example extraction recognizes it
only when it immediately follows an entry-point call. Functional and
example-preservation markers assign the corresponding labels to each description segment $d_i$.

We evaluate contract-only masking against reference segment labels
independently produced by GPT-5.4-mini from the problem specifications. Masking
achieves $95.48\%$ accuracy across 1,882 description segments, and $98.83\%$
of the removed segments are labeled contract-only. The remaining errors follow three patterns. 
(i) Generic references to arguments yield high semantic similarity but insufficient lexical overlap. 
(ii) Contract statements without a close counterpart in the extracted condition set fail both grounding criteria. 
(iii) Semantically aligned segments fall just below the similarity threshold despite strong lexical overlap. 
Segments with both a functional label and a contract-grounding edge are intentionally retained, protecting
information required for functional implementation while leaving part of the
contract-related content in the functional view.

We also evaluate executable-example extraction against reference line-level
labels independently produced by GPT-5.4-mini. The extractor achieves
$97.09\%$ accuracy, $98.19\%$ precision, $90.69\%$ recall, and an F1 score of
$94.29\%$. The remaining errors follow two patterns.
(i) \emph{Incomplete executable forms}: the example does not provide a
complete entry-point call and expected value in a single parseable unit, so
the extractor cannot convert it directly into an executable check. These
formats include doctest examples whose expected values appear on the following
line, natural-language statements without an explicit function call,
multi-line \texttt{Input:}/\texttt{Output:} pairs, and variable assignments
followed by separate results.
(ii) \emph{Symbolic recurrence forms}: recurrence relations use
call-to-expression notation, such as
\texttt{fib4(n) -> fib4(n-1) + fib4(n-2) + fib4(n-3) + fib4(n-4)}.
Although this notation resembles an input--output example, \texttt{n} is a
symbolic variable, so the extracted expression does not define a concrete
executable test case.

\begin{table}[H]
\centering
\footnotesize
\setlength{\tabcolsep}{6pt}
\renewcommand{\arraystretch}{1.15}
\begin{tabular}{lcp{0.58\textwidth}}
\toprule
\textbf{Set} & \textbf{Count} & \textbf{Markers} \\
\midrule
Functional
& 15
& \texttt{task}, \texttt{goal};\;
  \texttt{write}, \texttt{compute}, \texttt{find}, \texttt{separate},
  \texttt{return}, \texttt{returns};\;
  \texttt{rule}, \texttt{rules}, \texttt{valid if};\;
  \texttt{format}, \texttt{output};\;
  \texttt{example}, \texttt{examples} \\
\addlinespace
Example-preservation
& 7
& \texttt{==>}, \texttt{=>}, \texttt{$\rightarrow$}, \texttt{->},
  \texttt{==}, \texttt{=};\;
  \texttt{return(s)} \\
\bottomrule
\end{tabular}
\caption{Functional and example-preservation markers used to label
description segments during specification-graph construction.}
\label{tab:marker_sets}
\end{table}

\section{Executable Example Coverage}\label{ap:extract_example}
SLICE extracted at least one candidate check for 321 of the 340
tasks~(94.41\%). Checks from 319 tasks~(93.82\%) provided a usable execution
signal, while the checks extracted from two tasks~(0.59\%) failed for every
candidate because they contained unresolved variables. The remaining 19
tasks~(5.59\%) did not yield a check. When no usable execution signal was
available, all candidates received the same execution score and selection
proceeded through \texttt{diff-lp}. Generated input-validation assertions were
retained without example-based screening when no executable example input was
available.

Coverage alone does not guarantee that every extracted check provides a
discriminative execution signal. In two tasks, recurrence-style expressions
containing unresolved variables were interpreted as input--output examples.
The resulting checks failed for every candidate, so all candidates received
the same execution score and no candidate was favored. Candidate selection
therefore proceeded through \texttt{diff-lp}, following the same behavior used
when no usable executable check is available. The extraction procedure
obtained checks for 94.41\% of the tasks, while non-discriminative or
unavailable checks did not interrupt candidate selection because
\texttt{diff-lp} provided the fallback signal.

\section{Candidate Selection Behavior}\label{ap:selection_behavior}
Table~\ref{tab:selection_behavior} examines which signal determines the
selected function body. The executable checks produced a unique
highest-scoring candidate for only 1.47\%--6.76\% of tasks, with an average of
2.94\% across the four models. In 0.29\%--15.88\% of tasks, the tied
candidates shared every line after normalization, leaving no difference region
to score; this outcome accounts for 9.93\% of tasks on average.
Consequently, \texttt{diff-lp} determined the selected function body for
77.35\%--97.65\% of tasks, with an average of 87.13\%. Executable checks
therefore narrow the candidate set when they provide a discriminative signal,
whereas difference-region log probabilities provide the primary selection
criterion for the remaining candidates.

The final selections also show that the procedure does not collapse to the
greedily decoded candidate. A sampled candidate was selected for 111 of 340
tasks on Llama-3.1-8B, 188 tasks on Qwen3.5-9B, 117 tasks on Gemma-4-12B-it, and
168 tasks on GPT-5.4-nano. These counts correspond to non-greedy selection
rates of 32.65\%, 55.29\%, 34.41\%, and 49.41\%, respectively. Across the
four models, sampled candidates were selected in 584 of 1,360 model--task
evaluations, yielding an average rate of 42.94\%. The candidate pool therefore
changes the final function body in a substantial proportion of tasks rather
than supplying redundant alternatives to the greedy candidate.

The combination of the two analyses further shows that \texttt{diff-lp} does
not systematically return the greedily decoded candidate. Across the four
models, 1,185 of the 1,360 selections were determined by \texttt{diff-lp},
while only 175 belonged to the \emph{Unique score} or \emph{Identical}
categories. Even under the conservative assumption that all 175 selections
outside the \texttt{diff-lp} category chose a sampled candidate, at least 409
of the 584 non-greedy selections occurred among the cases resolved by
\texttt{diff-lp}. This lower bound corresponds to 70.03\% of all non-greedy
selections and 34.51\% of the selections determined by \texttt{diff-lp}.
Thus, difference-region scoring selects sampled implementations in a
substantial fraction of its decisions rather than serving as a nominal
tie-breaking rule that defaults to greedy decoding. This result is consistent
with the ablation study reported in the main text, in which replacing the
five-candidate generation and selection procedure with a single greedy body
reduced SSR and pass@1 by averages of 4.07\% and 4.36\%, respectively.

\begin{table}[H]
\centering
\small
\setlength{\tabcolsep}{3pt}
\renewcommand{\arraystretch}{0.95}
\begin{tabular*}{0.75\textwidth}{l@{\extracolsep{\fill}}ccc}
\toprule
\textbf{Model} & \textbf{Unique score} & \textbf{Identical} & \textbf{Tie-breaking} \\
\midrule
Llama-3.1-8B  & 6.76\% & 15.88\% & 77.35\% \\
Qwen3.5-9B    & 2.06\% & 0.29\% & 97.65\% \\
Gemma-4-12B-it   & 1.47\% & 13.24\% & 85.29\% \\
GPT-5.4-nano  & 1.47\% & 10.29\% & 88.24\% \\
\bottomrule
\end{tabular*}
\caption{Function-body selection outcomes. \emph{Unique score}: executable
checks select one candidate; \emph{Identical}: tied candidates have no
differing lines; \emph{Tie-breaking}: \texttt{diff-lp} selects among tied
candidates. Each row sums to 100\%.}
\label{tab:selection_behavior}
\end{table}

\section{Case Study: Complete Enforcement and Silent Acceptance} \label{ap:case_study}
We present a case study to examine why SLICE achieves higher contract
satisfaction than SpecFix and CodeTree without reducing functional correctness.
On HumanEval/121, the gold code and all three generated implementations achieved
100.00\% functional correctness on the valid-input tests, holding functional
performance constant across methods. Their behavior differed substantially on
the contract-violating tests~(CVTs). The gold code and SLICE intentionally
rejected 100.00\% of the CVTs through explicit assertions. CodeTree
intentionally rejected 80.00\% but failed to reject the remaining 20.00\%.
SpecFix contained no input-validation code, intentionally rejected 0.00\% of
the CVTs, and silently accepted 100.00\%. Among the three generated
implementations, only SLICE enforced all three stated contract conditions and
satisfied the task-level contract criterion. This comparison isolates contract
enforcement as the source of the performance difference.

Figure~\ref{fig:case_gold} presents the gold code, a human-written reference
implementation that explicitly enforces all three stated contract conditions.
The input must be a list, every element must be an integer, and the list must
contain at least one element. The implementation assigns one assertion to each
condition, producing a direct correspondence between the input contract and
its validation code. It returned correct outputs on all valid-input tests and
intentionally rejected all CVTs through \texttt{AssertionError}.

Figure~\ref{fig:case_slice} presents the function generated by SLICE.
It achieved 100.00\% condition coverage by representing all three stated
contract conditions through separate assertions.
The list-type assertion appears first, preventing a non-list input from reaching the subsequent
iteration. The element-type assertion then verifies every member of the list,
and the final assertion rejects an empty list. The final check is required
because \texttt{all(isinstance(x, int) for x in [])} evaluates to
\texttt{True}. SLICE achieved 100.00\% functional correctness on the
valid-input tests and intentionally rejected 100.00\% of the CVTs through
\texttt{AssertionError}. The generated implementation therefore achieved
complete contract enforcement without altering the required functional
behavior.

Figure~\ref{fig:case_specfix} presents the function generated by SpecFix.
It includes no input-validation code, resulting in 0.00\% condition coverage.
100.00\% of the CVTs proceeded through the function body and returned a value,
corresponding to a 100.00\% silent acceptance rate and a 0.00\% intentional
rejection rate. For example, \texttt{solution([3.0])} returned 3.0 because the
floating-point element passed through the arithmetic operations in the function
body. This behavior produced no exception or rejection signal indicating that
the input violated the contract. SpecFix therefore achieved 100.00\%
functional correctness on the valid-input tests but failed the task-level contract criterion.

Figure~\ref{fig:case_codetree} presents the function generated by CodeTree.
It achieved 33.33\% condition coverage by enforcing only one of the three
stated contract conditions.
CodeTree generated validation code only for the requirement that every list element be an integer, while
omitting the input-list type and non-empty-list requirements. The generated
check intentionally rejected 80.00\% of the CVTs through
\texttt{ValueError}. The remaining 20.00\%, corresponding to the empty-list
input, passed the check because
\texttt{all(isinstance(num, int) for num in lst)} evaluates to \texttt{True}
when \texttt{lst} is empty, after which \texttt{sum([])} returned 0. CodeTree
therefore achieved an 80.00\% intentional rejection rate and a 20.00\% silent
acceptance rate. Despite achieving 100.00\% functional correctness on the
valid-input tests, it failed the task-level contract criterion because CSR
requires 100.00\% of the CVTs to be intentionally rejected.

The case study isolates differences in contract enforcement while holding
functional correctness constant. All three generated implementations achieved
100.00\% functional correctness on the valid-input tests. SLICE achieved
100.00\% condition coverage and intentionally rejected 100.00\% of the CVTs.
CodeTree achieved 33.33\% condition coverage and intentionally rejected
80.00\% of the CVTs, while silently accepting the remaining 20.00\%. SpecFix
achieved 0.00\% condition coverage and intentionally rejected 0.00\% of the
CVTs, resulting in a 100.00\% silent acceptance rate. These results show that
the performance difference arises from contract enforcement rather than
functional implementation.

\begin{figure}[H]
\centering
\begin{codestyle}{Gold code}
def solution(lst):
    assert type(lst) == list, "invalid inputs" # $_CONTRACT_$
    assert all(type(x) == int for x in lst), "invalid inputs" # $_CONTRACT_$
    assert len(lst) > 0, "invalid inputs" # $_CONTRACT_$
    return sum(lst[i] for i in range(len(lst)) if i % 2 == 0 and lst[i] % 2 == 1)
\end{codestyle}
\caption{Gold code for HumanEval/121.}
\label{fig:case_gold}
\end{figure}

\begin{figure}[H]
\centering
\begin{codestyle}{SLICE~(ours)}
def solution(lst):
    assert isinstance(lst, list)
    assert all((isinstance(x, int) for x in lst))
    assert len(lst) > 0
    return sum((num for (i, num) in enumerate(lst) if i % 2 == 0 and num % 2 != 0))
\end{codestyle}
\caption{Function generated by SLICE with Llama-3.1-8B-Instruct for HumanEval/121.}
\label{fig:case_slice}
\end{figure}

\begin{figure}[H]
\centering
\begin{codestyle}{SpecFix}
def solution(lst):
    total_sum = 0
    for i, num in enumerate(lst):
        if i % 2 == 0 and num % 2 != 0:
            total_sum += num
    return total_sum
\end{codestyle}
\caption{Function generated by SpecFix with Llama-3.1-8B-Instruct for HumanEval/121.}
\label{fig:case_specfix}
\end{figure}

\begin{figure}[H]
\centering
\begin{codestyle}{CodeTree}
def solution(lst):
    if not all(isinstance(num, int) for num in lst):
        raise ValueError("The list must contain only integers.")
    return sum([num for i, num in enumerate(lst) if i % 2 == 0 and num % 2 != 0])
\end{codestyle}
\caption{Function generated by CodeTree with Llama-3.1-8B-Instruct for HumanEval/121.}
\label{fig:case_codetree}
\end{figure}

\section{Detailed Error Decomposition}\label{ap:detailed_error}
Table~\ref{tab:error} extends the main-text error analysis to all four
evaluated models. $\mathrm{F}^{+}$ indicates that the generated implementation
passes all valid tests after its input-validation code is removed, whereas
$\mathrm{F}^{-}$ indicates failure on at least one valid test.
$\mathrm{C}^{+}$ requires the input-validation code to reject every
contract-violating input without rejecting any valid input, and
$\mathrm{C}^{-}$ denotes failure of either requirement. Accordingly,
$\mathrm{F}^{+}\mathrm{C}^{+}$ equals SSR,
$\mathrm{F}^{-}\mathrm{C}^{+}$ denotes an incorrect function body with precise
contract enforcement, and $\mathrm{F}^{-}\mathrm{C}^{-}$ denotes failure in
both dimensions. $\mathrm{F}^{+}\mathrm{C}^{-}$ is classified as missed
contract violations if any contract-violating input is not rejected, and
otherwise as over-rejection if any valid input is rejected.

For SLICE, $\mathrm{F}^{-}\mathrm{C}^{+}$ is the largest failure category on
every model, accounting for 15.29\%--29.12\% of all tasks and
66.45\%--75.28\% of the tasks that do not achieve SSR. In contrast,
$\mathrm{F}^{+}\mathrm{C}^{-}$, obtained by combining over-rejection and
missed contract violations, accounts for 3.53\%--9.12\%, while
$\mathrm{F}^{-}\mathrm{C}^{-}$ accounts for 1.47\%--5.59\%. Averaged across
the four models, these three failure groups account for 20.74\%, 5.66\%, and
3.38\%, respectively. The dominant residual error under SLICE therefore lies
in functional implementation: the precise contract-enforcement criterion is frequently satisfied even when
the selected function body remains incorrect.

The comparison with prior methods shows that SLICE achieves the strongest
overall balance across the failure categories rather than minimizing a single
category. 
CodeTree records a slightly lower average
$\mathrm{F}^{-}\mathrm{C}^{+}$ rate of 19.41\%, compared with 20.74\% for
SLICE. However, relative to CodeTree, SLICE reduces
$\mathrm{F}^{+}\mathrm{C}^{-}$ by 49.69\%, from 11.25\% to 5.66\%, and
reduces $\mathrm{F}^{-}\mathrm{C}^{-}$ by 63.54\%, from 9.27\% to 3.38\%.
Its average missed contract violation rate is also 60.65\% lower, decreasing
from 6.91\% to 2.72\%. 
SpecFix records smaller
$\mathrm{F}^{-}\mathrm{C}^{+}$ and over-rejection regions, but misses contract
violations on 40.30\% of tasks and fails in both dimensions on 29.04\%.
Relative to SpecFix, SLICE reduces these two failure rates by 93.25\% and
88.36\%, respectively. Prompting methods exhibit a different imbalance, with
$\mathrm{F}^{-}\mathrm{C}^{+}$ reaching 39.12\% and over-rejection reaching
12.06\%. These comparisons show that SLICE substantially reduces contract-side
and joint failures while keeping functional failure under precise contract-enforcement success
comparable to CodeTree, resulting in the highest SSR across all four models.

The model-level distributions confirm that this advantage holds consistently
despite differences in the composition of the remaining errors. Under SLICE,
Llama-3.1-8B records the largest
$\mathrm{F}^{-}\mathrm{C}^{+}$ and aggregate
$\mathrm{F}^{+}\mathrm{C}^{-}$ rates, at 29.12\% and 9.12\%,
respectively. Qwen3.5-9B records 17.94\% and 3.53\%, while Gemma-4-12B-it
records 15.29\% and 4.71\%. GPT-5.4-nano records 20.59\% in
$\mathrm{F}^{-}\mathrm{C}^{+}$, 5.29\% in
$\mathrm{F}^{+}\mathrm{C}^{-}$, and 1.47\% in
$\mathrm{F}^{-}\mathrm{C}^{-}$. The internal composition of
$\mathrm{F}^{+}\mathrm{C}^{-}$ differs across models: missed contract
violations exceed over-rejection on Llama-3.1-8B and Gemma-4-12B-it, whereas
over-rejection is more frequent on Qwen3.5-9B and GPT-5.4-nano.
Nevertheless, SLICE keeps over-rejection between 2.06\% and 4.12\%, missed
contract violations between 1.47\% and 5.00\%, and joint failure between
1.47\% and 5.59\%. SLICE consequently achieves the lowest aggregate failure
rate, equivalently the highest SSR, on all four models. Relative to the
strongest competing method for each model, it reduces the aggregate failure
rate by 13.38\% on Llama-3.1-8B, 17.13\% on Qwen3.5-9B, 3.83\% on
Gemma-4-12B-it, and 8.83\% on GPT-5.4-nano, with an average reduction of
10.79\%. The full decomposition therefore confirms that specification-level
isolation provides the strongest overall balance between functional
implementation and contract enforcement. The remaining errors are concentrated
primarily in functional body generation, identifying the next bottleneck after
contract-side failures have been substantially reduced.

\begin{table}[t]
\centering
\footnotesize
\setlength{\tabcolsep}{4pt}
\renewcommand{\arraystretch}{0.92}
\begin{tabular*}{0.92\textwidth}{l@{\extracolsep{\fill}}lccccc}
\toprule
\multirow{2}{*}{\textbf{Model}} & \multirow{2}{*}{\textbf{Method}} & \multirow{2}{*}{$\mathrm{F}^{+}\mathrm{C}^{+}$~($\uparrow$)} & \multirow{2}{*}{$\mathrm{F}^{-}\mathrm{C}^{+}$~($\downarrow$)} & \multicolumn{2}{c}{$\mathrm{F}^{+}\mathrm{C}^{-}$~($\downarrow$)} & \multirow{2}{*}{$\mathrm{F}^{-}\mathrm{C}^{-}$~($\downarrow$)} \\
\cmidrule(lr){5-6}
& & & & over-rej. & missed & \\
\midrule
\multirow{7}{*}{Llama-3.1-8B}
& Base            & 46.18\% & 30.00\% & 9.12\% & 2.35\% & 12.35\% \\
& Few-shot        & 44.41\% & 34.71\% & 10.29\% & 0.88\% & 9.71\% \\
& CoT             & 43.82\% & 29.12\% & 12.06\% & 2.06\% & 12.94\% \\
& Self-planning   & 43.24\% & 39.12\% & 5.88\% & 2.06\% & 9.71\% \\
& CodeTree        & 49.41\% & 29.71\% & 1.76\% & 10.59\% & 8.53\% \\
& SpecFix         & 18.53\% & 8.82\% & 0.29\% & 32.35\% & 40.00\% \\
& \textbf{SLICE~(ours)} & 56.18\% & 29.12\% & 4.12\% & 5.00\% & 5.59\% \\
\midrule
\multirow{7}{*}{Qwen3.5-9B}
& Base            & 66.76\% & 25.00\% & 3.82\% & 1.47\% & 2.94\% \\
& Few-shot        & 59.41\% & 26.76\% & 9.12\% & 0.29\% & 4.41\% \\
& CoT             & 69.12\% & 20.00\% & 4.41\% & 2.35\% & 4.12\% \\
& Self-planning   & 66.76\% & 23.82\% & 3.82\% & 1.76\% & 3.82\% \\
& CodeTree        & 55.29\% & 11.76\% & 3.24\% & 6.47\% & 23.24\% \\
& SpecFix         & 5.29\% & 0.59\% & 0.59\% & 60.59\% & 32.94\% \\
& \textbf{SLICE~(ours)} & 74.41\% & 17.94\% & 2.06\% & 1.47\% & 4.12\% \\
\midrule
\multirow{7}{*}{Gemma-4-12B-it}
& Base            & 76.76\% & 16.47\% & 2.65\% & 2.35\% & 1.76\% \\
& Few-shot        & 68.24\% & 17.94\% & 10.88\% & 0.29\% & 2.65\% \\
& CoT             & 75.59\% & 17.35\% & 2.65\% & 2.35\% & 2.06\% \\
& Self-planning   & 71.18\% & 21.18\% & 2.35\% & 3.53\% & 1.76\% \\
& CodeTree        & 71.76\% & 14.12\% & 2.65\% & 8.24\% & 3.24\% \\
& SpecFix         & 28.82\% & 4.12\% & 0.59\% & 45.59\% & 20.88\% \\
& \textbf{SLICE~(ours)} & 77.65\% & 15.29\% & 2.06\% & 2.65\% & 2.35\% \\
\midrule
\multirow{7}{*}{GPT-5.4-nano}
& Base            & 70.00\% & 23.53\% & 4.41\% & 0.29\% & 1.76\% \\
& Few-shot        & 64.12\% & 24.12\% & 9.12\% & 0.00\% & 2.65\% \\
& CoT             & 70.00\% & 22.65\% & 5.29\% & 1.18\% & 0.88\% \\
& Self-planning   & 67.65\% & 22.94\% & 4.41\% & 2.35\% & 2.65\% \\
& CodeTree        & 63.82\% & 22.06\% & 9.71\% & 2.35\% & 2.06\% \\
& SpecFix         & 45.00\% & 9.12\% & 0.88\% & 22.65\% & 22.35\% \\
& \textbf{SLICE~(ours)} & 72.65\% & 20.59\% & 3.53\% & 1.76\% & 1.47\% \\
\bottomrule
\end{tabular*}
\caption{Error decomposition. $\mathrm{F}^{+}$ and $\mathrm{F}^{-}$ indicate
whether the generated implementation passes or fails the valid tests after
input-validation code is removed. $\mathrm{C}^{+}$ requires the
input-validation code to intentionally reject every contract-violating input
without rejecting any valid input, whereas $\mathrm{C}^{-}$ denotes failure
of either requirement. Thus, $\mathrm{F}^{+}\mathrm{C}^{+}$ equals SSR.}
\label{tab:error}
\end{table}

\end{document}